# On Cloud-based Oversubscription

Rachel Householder[#1], Scott Arnold[#2], Robert Green[#3*]

*#Computer Science Department, Bowling Green State University, USA*

*Abstract—* **Rising trends in the number of customers turning to the cloud for their computing needs has made effective resource allocation imperative for cloud service providers. In order to maximize profits and reduce waste, providers have started to explore the benefits of oversubscribing cloud resources. However, the benefits of oversubscription in the cloud are not without inherent risks. In this paper, we attempt to unveil the different incentives, risks, and techniques behind oversubscription in a cloud infrastructure. Additionally, we provide an overview of research that has been completed on the topic and make suggestions for the direction of future work.**

## I. INTRODUCTION

Utilizing cloud services to meet computing needs is a concept that is rapidly gaining in popularity as "in the cloud" has become a catchphrase in mainstream society. According to NIST, *"Cloud computing is a model for enabling convenient, on-demand access to a shared pool of configurable computing resources (e.g. networks, servers, storage applications, and services) that can be rapidly provisioned and released with minimal management effort or service provider interaction"* [1]. The resources offered by cloud service providers (known from here on out as CSPs) can be classified under one of three service models. Software as a Service (SaaS) provides customers with access to applications that run on the foundation of a cloud infrastructure. Platform as a Service (PaaS) provides the tools to create and/or implement applications on top of a cloud infrastructure with customer control being limited to the environment of these applications. On the lowest level, Infrastructure as a Service (IaaS) provides access to resources such as servers, storage, hardware, operating systems, and networking. Unlike SaaS and PaaS, the customer has the ability to configure these lower-level resources.

IaaS has become increasingly popular [21] as it allows customers, especially companies and organizations, to outsource their IT needs. These companies simply request the computing resources they desire [11-12] and CSPs provide those resources with a high level of assurance of their reliability and availability. The outsourcing of computing resources has several benefits for customers. Services are offered on a pay-as-you-go basis, allowing customers to only pay for the resources they use. CSPs handle much of the IT infrastructure management tasks that customers once had to support themselves. Additionally, data and services in the cloud are widely available through the internet and can be accessed using a variety of devices.

With all of these benefits, the number of customers looking to migrate to the cloud is on the rise and the ability of CSPs to efficiently host as many clients as possible on a fixed set of physical assets will be crucial to the future success of their business [2]. Cloud services are supplied to clients through virtualization creating the impression that each user has full access to a seemingly unlimited supply of resources. In reality, a single physical machine must divide its finite set of resources amongst multiple virtual machines (VMs). Much research has been dedicated to developing optimum resource allocation strategies in a non-overbooked cloud. For instance, [3] uses concepts of Queuing Theory to maximize revenues and increase resource utilization levels while adhering to Service Level Agreement (SLA) constraints. [4] employs a multivariate probabilistic model to optimize resource allocation.

While these strategies have been shown to improve utilization, a high percentage of resources still sit idle at any given time [5]. As a result, oversubscription of cloud services has become an appealing solution to further optimize cloud efficiency. This paper discusses the concept of oversubscription in the cloud and the work that has been done towards making it a viable solution for reducing wasted resources. The rest of the paper is organized as follows:

- Section II: We provide an overview of the concept of oversubscription.
- Section III: We share a literature review of the work that has been done on the topic.
- Section IV: We provide insights into the future direction of uses and research on the topic.
- Section V: We conclude with our final thoughts.

## II. OVERVIEW OF OVERSUBSCRIPTION

### A. Oversubscription In Other Industries

To oversubscribe a resource means to offer more of that resource than there is actually capacity for under the assumption most customers will not actually consume their entire portion. The goal is to diminish the sum of unutilized resources and thus increase profits.

Oversubscribing resources to reduce waste and maximize profits is not a concept unique to cloud computing. Hotels overbook rooms with the assumption that not all clients will show up for their stay. However, as Noone and Lee [6] explain, sometimes there are less no-shows than expected, resulting in some customers having to downgrade to a lower standard of room or find another hotel in which to stay. To limit customer dissatisfaction for the denial of services, customers are typically given monetary compensation or a voucher. However, this does not always placate customers enough to retain them for future stays.

The healthcare industry also uses overbooking strategies to schedule patient appointments [7]. As in the lodging industry, doctors experience a waste of resources, specifically the doctor's time, when patients do not show up for appointments. However, booking too many patients can lead to an increase in patient waiting times and physician overtime. Patient characteristics such as history of no-shows can be taken into consideration in order to determine which time slots can be overbooked.

The airline industry is another commonly known for overbooking resources. Airlines have been known to overbook seats [8] and cargo space [9]. When more customers show up than predicted, they are typically moved to another flight which causes delays and other inconveniences for the customer. To better accommodate passengers when this occurs, airlines will sometimes form alliances with their competitors to expand the number of flights bumped customers can be moved to [10]. The impact of this alliance can be taken into consideration when developing overbooking policies. Additionally, the class system can be taken into consideration when developing overbooking policies with first-class flights typically having lower levels of overbooking than coach [8].

### B. Oversubscription in Cloud Computing

Like the lodging, healthcare, and airline industries, cloud computing provides ample opportunity for oversubscription. In recent years, companies have started to notice that they are only utilizing a small portion of their available resources (resources being memory, CPU, disk, and bandwidth capacity). In fact, CSPs on average use only 53% of the available memory, while CPU utilization is normally only at 40% in most data centers [5]. Simulations done to study the CPU utilization patterns of individual VMs have shown that 84% of VMs reach their maximum utilization levels less than 20% of the time. The results of this study performed by Ghosh and Naik [11-12] are illustrated in the histogram pictured in Figure 1. The underutilization of resources is a major concern to most CSPs considering the amount of resources required to run and maintain large data centers. Data centers require a great deal of infrastructure that consumes large amounts of power [1]. Oversubscription helps to maximize resource utilization which can in-turn help to reduce these costs and increase profitability.

In cloud computing, a cloud is said to be oversubscribed when the sum of customers' requests for a resource exceeds the actual physical available capacity. There can be oversubscription on both the customer's end and the provider's end [2]. Oversubscription stemming from the customer occurs when they do not reserve enough computing power to meet their needs. Oversubscription on the provider's end occurs when CSPs book more requested capacity than they can actually support. This type of oversubscription is more common than the former as many customers actually tend to reserve more resources than they need [11-12]. Thus, we will focus on overbooking by the CSP in this paper.

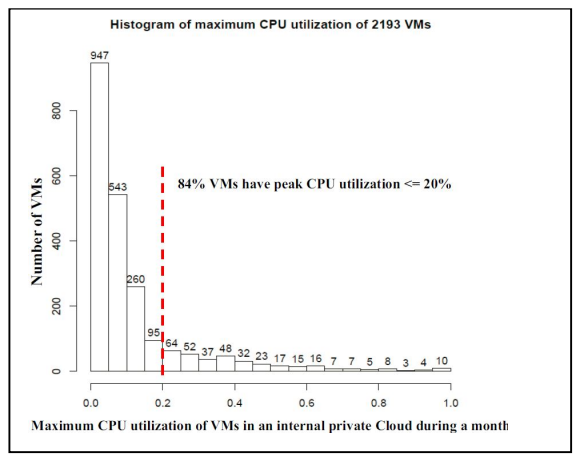

Fig. 1. Shows most VMs rarely have peaks CPU utilization [11-12].

Figure 2 shows an example of a non-oversubscribed cloud using memory capacity as an example. Each column represents a VM sitting on top of the physical machine (PM). The bottom half is the actual amount of memory being utilized by the customer while the top half is the space left unused. In this model, the sum of memory allotted to all VMs does not exceed the actual capacity of the PM. Here 37.5% of the available physical memory is being utilized. Conversely, Figure 3 shows a corresponding example for an oversubscribed PM. Six VMs that have reserved a total of 6GB of memory are allocated on a PM with a physical capacity of 4GB. The model increases resource utilization to 68.75%.

## C. Risks of Cloud Oversubscription

Though there are valid motivations to oversubscribe computing resources, the strategy is not without inherent risks. In order to make oversubscription possible, providers must make several assumptions about their customers. They assume that not all customers will use any or all of their requested resources. They also assume that not all customers will show up to use their resources at the same time. By making these assumptions, CSPs flirt with the possibility of running out of resources that customers have legitimately paid for. This can have costly consequences for CSPs, one of which is overload.

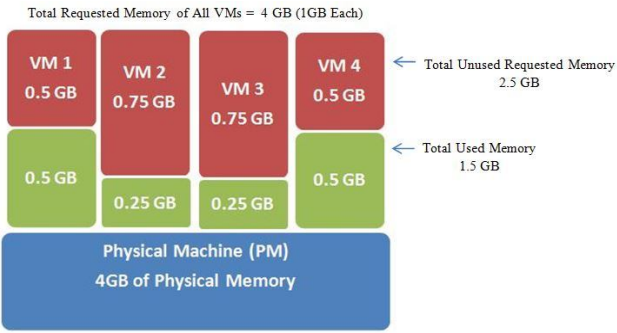

Fig. 2. VM memory allocation on an oversubscribed PM

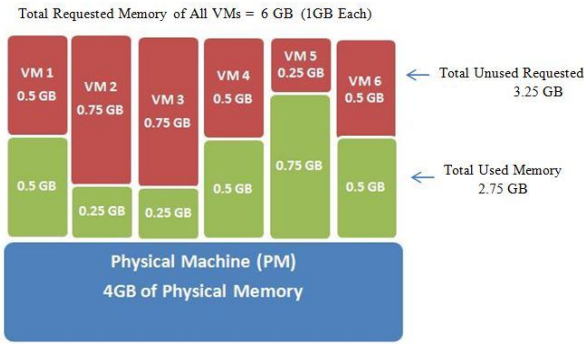

Fig. 3. VM memory allocation on an oversubscribed PM

Overload occurs when the infrastructure is strained to meet demands as requests for resources near or exceed the physical capacity. This can severely degrade the performance of the cloud and even lead to outages and failures for some customers [1-2, 5, 11-23]. CPU overload can result in an abundance of processes waiting to run in a VMs CPU queue [13-14]. This can degrade application performance and reduce the ability of cloud monitoring agents to supervise VMs. Disk overload, which is not thoroughly discussed in this paper, can have similar consequences in regards to performance reduction.

If memory becomes overloaded (usually classified by multiple VMs swapping out to the disk) it can be devastating because it has the potential to inhibit any application progress [2]. Large overheads and thrashing can seriously impede the performance of the system [2, 13-14].

Finally, network overload can cause bottlenecks that slow progress and can lead to a reduction in resource utilization and oversubscription gains [13-14, 17, 19-22].

If overload is not managed, CSPs further run the risk of violating its Service Level Agreements (SLAs). SLAs provide customers with a sense of security by providing a level of assurance that their requested resources will be available and operational when they need them [11-12]. Some SLAs are legally bonded, which means that companies could be forced to provide compensations to customers for SLA violations. Even one SLA violation can be costly to CSPs and so developing an oversubscription policy that considers SLA constraints is crucial for effective implementation.

## D. Overload Prevention and Mitigation

In developing a model for oversubscription, CSPs must take both proactive and reactive steps to reduce overload. Studying client habits is one predictive measure taken to determine how best to allocate resources so as to optimize oversubscription while reducing performance degradation from overload [1-2, 5, 11-12, 16, 18, 21, 23].

Even with proactive preventative resource allocation models in place, overload of resources can still occur. When it does, strategies to effectively detect and manage overload must be employed. A basic description of some common overload remediation techniques are discussed by Baset et al. [13-14] and are as follows:

- Stealing is the act of taking resources from under loaded VMs and giving them to overloaded VMs on the same physical host. Memory ballooning is a common example of this and it is often a capability installed in hypervisors that use this as a first line of defense against overload.

- Quiescing [13-14, 18] is the act of shutting down VMs on an overloaded PM so that the remaining VMs can function at normal performance levels. The VMs that are shut down will resume once overload subsides.

- VM Migration [1-2, 13-15, 17-21] is the act of moving a VM from one physical machine to another. This is typically done in systems that have VM storage located on Storage Area Network (SAN) devices as opposed to local memory. In the former situation, only a memory footprint of a VM needs to be moved while in the latter, VM migration to the disk can encumber the data center network.

- Streaming disks serves as a solution to reduce the network costs of migration to the disk. Here, a portion of the VMs disk is transferred to another PM. When the transfer is completed, the migrated VM can access disk space on both the original and new PM. Once the network traffic is low, the two disks can be reunited.

- Network memory can reduce load caused by swapping on the PM disk by allowing the memory of another PM to be used for swapping over the network.

Research on the utilization of some of these techniques is discussed further in the literature review.

### III. LITERATURE REVIEW

#### A. Memory Oversubscription

As discussed previously, there are several resources offered by CSPs that are candidates for overbooking including memory, disk, CPU, and bandwidth. The ability to effectively oversubscribe memory in particular provides a challenging problem for researchers [16]. In their overview of oversubscription, Wang et al. [13-14] test the ability of quiescing and live migration to remediate overload in a memory oversubscribed cloud. Their experiments show that quiescing can be used successfully to lower the number of VMs experiencing overload. Additionally, live migration can also remedy overloads. However, due to network constraints on the number of VMs that can be migrated in a given time period, live migration may not be effective as a sole method. Utilizing a combination of quiescing and live migration techniques is suggested as a possible improvement of the study.

As a continuation of the latter work, Wang, Hosn, and Tang [18] introduce the *ROWM* algorithm that utilizes the concept of a work value as the driving factor in memory overload remediation. *ROWM* is an acronym for Remediating Overload with Work Value Maximized. This study assumes that resources are not guaranteed to be available for migration due to the oversubscribed nature of the system. When this is the case during overload, migration is not an option. Thus, quiescing and resuming VMs are used in conjunction with migration. To determine which VMs to quiesce or resume, each VM is assigned a work value based on the expected work to be done and the impact on other VMs. VMs with lower work values are chosen first to be quiesced in order to allow those with higher work values to continue normally. When compared to a baseline algorithm, tests show a work value increase of 27% in oversubscribed settings for the *ROWM* algorithm. Additionally, overall service performance improves by 40%.

Williams et al. [2] present *Overdriver* in an attempt to reduce or eliminate performance degradation that comes with memory overload. *Overdriver* is a flexible system that classifies the type of overload as either transient or sustained based on a threshold of duration set by the VMs probability overload profile. The type distinction is then used to determine the appropriate overload mitigation strategy to implement. Transient overloads which are most common are shorter in duration. *Overdriver* handles overloads of this type using a new method called cooperative swap which takes swap pages from overloaded VMs and stores them in repositories elsewhere on the network. For longer, sustained overloads, VM migration is used for remediation. The goal is to alleviate the high overhead associated with VM migration by using this method only when necessary for sustained overloads. Evaluation shows that *Overdriver's* throughput is within 8% of a non-oversubscribed VM. Additionally, *Overdriver* has less overhead than an approach that solely uses VM migration.

Hines et al. [16] also undertake the problem of managing resource allocation in a memory oversubscribed cloud. They present a framework called *Ginkgo* that can be used to automate the redistribution of memory among VMs. *Ginkgo* uses application performance as the criteria for memory allocation decisions. To do this, *Ginkgo* first builds a performance profile for each VM. The performance profile, along with other constraints such as performance level thresholds established by the SLA, is used to determine the amount of memory each VM requires. The memory is then allocated to each VM and the process is repeated. To further improve performance, a variation of ballooning called JavaBalloon can be enabled to dynamically adjust the heap size of a Java Virtual Machine (JVM). Testing applications of the *Ginkgo* framework show significant savings in physical memory (up to 73%) with minimal negative effects on performance (within 7% degradation when compared to a non-oversubscribed system). This essentially results in the ability to host more VMs in the system.

#### B. Bandwidth Oversubscription

Bandwidth is another resource that can be oversubscribed. As the cloud gains in popularity, network traffic continues to increase which can lead to bottlenecks and latency [19-20]. Jain et al. [17] focus on remediating overload using VM migration in a tree topology data center network. They are the first to concurrently consider load constraints of servers and the traffic capacity constraints of the tree edges. They characterize each VM using the following metrics: 1) transfer size, 2) computational load, and 3) value of migration, and they use these metrics to determine which VMs are good candidates for migration. Next, they define which servers are "hot" and which servers are "cold". The ultimate goal is to relieve as many hot servers as the network will allow by migrating a portion of their VMs to cold servers. This is a version of the constrained migration problem (CoMP). To determine where VMs should be migrated to, the topology of the tree is considered. The shorter the path the VM must take to get to its new destination, the less costs incurred through migration. Three algorithms are proposed: 1) *Single hot server* selects VMs for migration. 2) *Maximum throughput* uses an undirected tree approach to maximize the number of VMs that can be migrated from hot servers. 3) *Multiple hot servers* uses a directed tree approach to maximize the number of hot servers that can be relieved. Their system called *WAVE* is shown in testing to relieve up to 67% of servers.

Guo et al. [22] focus on optimizing the distribution of network traffic and throughput using a load balancing technique in a fat-tree data center network. The fat-tree data center topology is used by many data center networks that support high performance distributed computing applications such as MapReduce. These applications require extensive communication between servers to process large data sets. In order to understand the possible risk of network bottlenecks and develop a multicast flow scheduling model, they analyze the "*minimum link oversubscription upper bound*". This analysis contributes to the development of their Oversubscription Bounded Multicast Scheduling (OBMS) algorithm. Experiments applying OBMS to a variety of traffic patterns demonstrate enhanced load balancing and throughput capabilities.

When the focus is on consolidating VMs so that they fit on a minimal number of physical hosts or racks, network congestion can become particularly cumbersome to performance as bottlenecks are more likely to occur. Because bandwidth usage tends to vary, estimating the demand can be difficult. As a result, resources are typically allocated so that the aggregate use of historical bandwidth for all VMs on a physical host does not exceed the network capacity. This practice proves to be wasteful. Breitgand and Epstein [19-20] attempt to improve bandwidth utilization by applying concepts of the Stochastic Bin Packing problem (SBP). The goal is to allocate VMs belonging to a single SLA class in such way that the probability of meeting bandwidth demands is at least the minimum calculated value. They suggest and test three algorithms: 1) *First Fit Variance Mean Ratio (VMR) Decreasing,* 2) *Fractional Algorithm,* and 3) *On-Line Algorithm.* They test these algorithms by comparing them to algorithms in previous works. Algorithm 2 is compared to First Fit (FF), First Fit Descending (FFD), and Group Packing (GP) and proves optimal for both real and synthetic data. Algorithm 1 outperforms FFD and Algorithm 3 proves superior to FF and GP.

Wo et al. [21] also focus on allocating VMs efficiently to reduce energy consumption. Like [22], they look at cloud data networks that support communication intensive applications like MapReduce. They use a greedy-based VM placement algorithm called *GreedySelePod* that is traffic-aware to improve the locality of servers running the same application. The algorithm shows improved application acceptance rates and overall performance. Additionally, they introduce a revenue model designed to determine the overbooking ratio that will maximize profits by reducing the number of SLA violations. Analysis of their model shows that when using the optimum overbooking ratio, revenue can be increased by 84.2%.

### C. CPU Oversubscription

Zhang et al. [15] consider a cloud that has overcommitted its processor power. They introduce a VM migration algorithm called *Scattered* that focuses on pinpointing the best VMs for migration based on evaluation of their workload degree of correlation. Once VMs are identified, the algorithm selects the optimal PMs to migrate to. The goal is to minimize the overall effect on the data center network by migrating as few VMs as possible for overload remediation. Their algorithm allows for two variations of migration. The first is standard migration where a VM is moved from an overloaded PM to an underloaded one. The second is an exchange of VMs where two VMs essentially swap places. The latter approach is practiced when a VM cannot fully fit on another machine, but the level of overload can still be reduced by swapping.

They compare their algorithm to the Largest VM First (LVF) migration algorithm. Scattered is shown to limit the number of migrations required to relieve overload and can tolerate larger overcommit ratios.

### D. Energy Efficiency

Overcommiting resources has the ability to allow more VMs to operate on a smaller number of PMs, meaning that more customers can be served using a smaller portion of physical resources. This is beneficial for CSPs because it reduces both energy and hardware costs. Moreno and Xu [1] focus on the value of overbooking for greener computing. Using a multi-layer Neural Network, they attempt to predict resource usage patterns by studying historical data for the current workload. This, along with the Optimal Allocation Limit (OAL), is used to define an over-allocation algorithm that identifies how to under-allocate resources to each VM, thus overbooking the total available resources. When overload does occur, they have a compensation policy in place to remediate the load. This is done by migrating VMs using a Largest Allocated First selection process based on CPU utilization. Simulations indicate that their approach improves energy efficiency and reduces the number of compensation events, deadline violations, and exceeded violation time.

### E. Optimum Resource Allocation

Ghosh and Naik [11-12] study the practice of CPU overbooking, but they do so by applying predictive risk analysis. They evaluate the history of CPU usage of VMs to establish a one-sided tolerance limit which represents a threshold of risk for overcommiting a set of VMs. This risk is based on the probability of overload and SLA violations. They propose that these analytics can be applied to develop a smart risk-aware resource allocation tool that can be used to place incoming requests.

Tomas and Tordsson [5] propose a whole framework for VM placement. An admission control module determines if a new job request can be deployed. Applications are monitored and profiled to help predict their behavior and predominant type of resource usage. This allows for greater flexibility so that applications with higher memory needs can be treated differently from applications with bursty CPU or I/O usage. A smart overbooking scheduler then determines the best location for the application to be deployed. Simulations performed indicate that resource utilization can be improved by 40% with this approach and the number of accepted applications can be multiplied by a factor of three.

Breitgand et al. [23] approach optimizing resource allocation by focusing on the structure of SLAs. Commonly, availability SLAs specify a probability that defines how often VMs will be available over a given time period. These SLAs do not provide a guarantee that a VM will be successfully launched. Currently, in the situation where a VM is not launched successfully, the provider makes it look as though it had been launched but then became unavailable. To improve transparency to the customer, a new extended SLA (eSLA) is suggested. The eSLA provides a probability for successful VM launch in addition to an availability probability. Using the eSLA as a constraint, a risk-aware model for VM placement is developed. The model is shown to allow for more aggressive overcommit ratios without increasing the number of eSLA violations.

## IV. DISCUSSION OF FUTURE WORK

Overbooking resources in the cloud infrastructure is a fairly new area of research, and thus there is much work that can be done to improve efficient resource allocation and management in an overbooked cloud. In the search to define best practices, algorithms for quantifying the risk of overbooking will likely continue to be studied as a means to improve resource utilization with minimal consequences. In terms of overload mitigation, strategies for selecting VMs to be reallocated, determining the appropriate method for moving selected VMs, and identifying the optimum destination for reallocated VMs will continue to be refined.

Much of the research that has been conducted thus far has focused on managing oversubscription of a single resource such as bandwidth, CPU, or memory. While this research is useful in progressing the efficiency and understanding of oversubscription techniques, the narrow focus is somewhat limiting. Future work that extends approaches for load balancing to multiple resource types will likely increase cloud providers' ability to efficiently host more virtual machines on a fixed set of physical resources.

Research on overbooking has also been predominantly focused on the IaaS model. IaaS provides a natural environment for overbooking because it is at this level that users can request access to basic resources such as storage, network, and computing power. These requests are granted through a veil of virtualization that allows for the possibility of oversubscribing resources unbeknownst to users. While overbooking is primarily implemented at the IaaS level, as [13-14] points out, it is possible for the concept of overbooking to be applied to both PaaS and SaaS. Future work on cloud oversubscription may branch out to explore these possibilities.

With the inherent risks of overbooking, violations of traditional SLAs may become more frequent [1]. As discussed in the literature review portion of this paper, several researchers have integrated SLA constraints into their models [1, 3, 5, 11-14, 16, 21-23] and the eSLA [23] has been suggested as a solution for oversubscribed clouds. With the trend going towards overbooking, further research should be done on how to define SLAs for an oversubscribed cloud environment.

There are noticeable similarities between overbooking in other industries and the cloud. For instance, overbooking too many appointments in a doctor's office can lead to longer wait times for patients and more time needed for the doctor to complete the overall job of seeing all waiting patients. This is analogous to how overbooking a resource such as CPU can potentially lead to longer waiting times if overload occurs. Additionally, airlines often have class systems that outline the level of services a passenger is entitled to. First class is more expensive than coach, but these passengers also receive a more luxurious flight. Similarly, some clouds use levels of service to offer customers flexibility in selecting a price for the computing experience they need [13-14]. In all industries, waiting times or lack of service availability is a potential consequence that can lead to customer dissatisfaction.

Researching the application of oversubscription in other industries raises new questions that can be addressed in future work on the cloud. Some of these questions are listed below:

- How much should customers be compensated in the event of an SLA violation to best maintain retention rates?

- Is one type of compensation (i.e. offering vouchers for computing resources as opposed to monetary compensation) better able to placate angry customers?

- Similar to the airline practices, could CSPs benefit from alliances with competitors by assigning workloads to them in times of overload? If so, what impact would that have on overbooking models when compared to more traditional models?

- Can having levels of service similar to airline classes improve overbooking policies by implementing varying degrees of overbooking to the different levels?

The answers to these questions may give CSPs an edge in the highly competitive market that is outsourcing computing power, yet there are many ways in which cloud-based oversubscription differs from other industries. Other industries tend to maximize oversubscription by optimizing a single available resource. This main available resource is what generates profit for the particular company, which is why utilization is so important. In comparison, CSP providers oversubscribe several resources with hopes to optimize the use of existing computing infrastructures. When these oversubscription methods fail, the results may be felt by all the subscribers of the service. Outside of the cloud, the impact of oversubscription within industries isn't always universal. Oversubscription can certainly fail, but it doesn't hold the reciprocal effect that it would in the cloud. For this reason, proper optimization of resources within the cloud is extremely important as it allows providers to offer competitive pricing and also ensures customer satisfaction.

Another topic that has not yet been addressed in the area of cloud-based oversubscription is the issue of redundancy, particularly geo-redundancy. When implementing geo-redundancy, the main goal typically falls into one of two categories: enhancing data durability or protecting service performance. Data durability, though of great importance in any type of networked, remotely accessed environment, can readily be solved through matters of replication and mirroring. Continued service performance, on the other hand, is an objective that can be in direct opposition to oversubscription. For instance, in applications of geo-redundancy it is typical for service to failover (switch to another data center) if a major failure occurs. Yet, if both data centers are currently oversubscribed, there may not be enough utilized resources to maintain performance. This results in either 1) Service degradation across the board (i.e. an additional application of oversubscription causing pain to all in order to continue serving all) or 2) Service failure (as there are no resources available to serve the new need, service provision is simply stopped). Applying this situation to oversubscription could lead to enhancements in algorithms for the design and evaluation of cloud computing systems, particularly those that are oversubscribed and must maintain a very high level of availability.

## V. CONCLUSION

Resource allocation models in cloud computing infrastructures tend to allow large fractions of resources to sit idle at any given time. The ability to improve resource utilization and decrease waste can significantly increase the profits of a CSP. In this paper, we discuss the application of oversubscription techniques to achieve this goal. We define overbooking in the context of the cloud and discuss the types of computing resources that can be overbooked. Overload is introduced as a risk of over-allocating resources and some common overload mitigation strategies are defined. Finally, we conduct a literature review to outline the state of the art, and we provide insights into the direction of future research on oversubscription.